\documentclass[12pt,reqno]{amsart}

\usepackage{lmodern}
\usepackage[T1]{fontenc}
\usepackage[utf8]{inputenc}

\usepackage{textcomp}

\usepackage{microtype}

\usepackage{mathtools}
\usepackage{amssymb}
\usepackage{amsxtra} 
\usepackage{amsthm}

\usepackage[mathscr]{eucal}    

\usepackage{bm}		

\usepackage{graphicx}

\usepackage{cite}

\usepackage{enumitem}

\usepackage{booktabs}

\usepackage{xcolor}

\usepackage{csquotes}

\usepackage{threeparttable}

\usepackage{xspace}

\usepackage{xfrac}

\usepackage{url}
\usepackage{siunitx}      

\theoremstyle{plain}

\newtheorem*{definition}{Definition}

\DeclareFontFamily{OT1}{pzc}{}
\DeclareFontShape{OT1}{pzc}{m}{it}{<-> s * [1.100] pzcmi7t}{}
\DeclareMathAlphabet{\gyre}{OT1}{pzc}{m}{it}

\DeclareMathAlphabet{\gk}{OML}{txmi}{m}{it}

\newcommand{\eu}{\ensuremath{\boldsymbol{\mathrm{e}}}}

\newcommand\eg{e.g.,\xspace}

\begin{document}


%
\title[Odds Ratios]{A NEW LOOK  AT THE ODDS RATIO \\ IN LOGISTIC REGRESSION}
\author{JOS\'E~RA\'{U}L~MART\'{I}NEZ} 
\email{j.r.martinez@mac.com}
\urladdr{https://www.researchgate.net/profile/Jose-Martinez}
\date{ Revised and expanded: May 2025. Original: December 2021}

\begin{abstract}

The standard odds ratio of logistic regression is foundational but limited to individual explanatory variables. This work derives a  multivariable odds ratio that applies to all the explanatory variables in all their combinations.

\end{abstract}


\maketitle

\newpage

\tableofcontents

\newpage
\section{Introduction} \label{sec:intro}
%

In a nutshell, this work is about extracting more information from the standard  odds ratios of  multivariable logistic regression models whose explanatory variables are binary and have no interactions. For such models, the standard odds ratios by themselves provide incomplete information because each applies only to an  individual explanatory variable. In this work, I derive a generalized multivariable odds ratio that treats all the variables jointly in all their combinations. The resulting ensemble of  odds ratios tells  the complete  odds-ratio story.

This document's  organization is as follows. Section \ref{sec:prelims} sets the stage   by defining  variables, notation, a general binary logistic regression model, and certain    mathematical objects that I shall call ``events.''   Sections \ref{sec:oddsratio}--\ref{sec:groupoddsratio} contain the main results. In Section \ref{sec:oddsratio},  I derive, analyze,  and interpret the known  standard odds-ratio formula for individual variables  with a view toward  recasting it. Recasting is done in Section \ref{sec:reformulation}, in which I derive and validate a new formula for the odds ratio of individual variables. In Section \ref{sec:groupoddsratio}, I derive a formula for a  generalized multivariable odds ratio and show  that the formula produces the entire ensemble of odds ratios.  Section \ref{sec:conclusions} contains  conclusions.

 \section{Preliminaries }      \label{sec:prelims}
%
 \subsection{General Binary Logistic Regression Model }      \label{subsec:themodel}
%
The variables and associated quantities used in this work are    
\begin{align}
 	y &  = \text{Binary response variable, coded } (0,1)  \label{eq:respvar}, \\ 
	x_n & = n\text{th binary explanatory variable, coded }(0,1), \label{eq:expvar}  \\
	n &= 1,2,\ldots,N, \label{eq:n} \\
	N & = \text{Number of binary explanatory variables}, \label{eq:N} \\
	X_N & = \{ x_1,x_2,\dots,x_N\} = \text{Set  of } N \text{ explanatory variables}. \label{eq:X}  
\end{align}
Hereafter, I shall  use ``variable(s)'' as shorthand for ``explanatory variable(s)'' unless otherwise indicated. All the variables are binary and noninteracting.

The general binary linear logistic regression model is known as the logit model or \textbf{logit} for short. The logit  version I use in this work is common in the literature (for example, see  \cite{hosmer}) and is defined in \eqref{eq:gxlogit} using the  notation  in \eqref{eq:respvar}--\eqref{eq:X}. 
\begin{align} 
	\boldsymbol{g}(X_N) = \log\big(\mathscr{O}(X_N)   \big)  &= \beta_0 + 	\beta_1 x_{1} + \beta_2 x_{2} + \cdots + \beta_N x_{N}, \label{eq:gxlogit} 
\end{align} 
where ``$\log(\cdot)$'' denotes the \emph{natural logarithm} function and 
\begin{align}
\begin{split} \label{eq:odds}
	\mathscr{O}(X_N) & = \frac{p \left( y  | X_N \right)  }{1 - p\left(y | X_N \right)   }, \\
		&= \text{ Odds of } y \text{ occurring given } X_N, \\
\end{split} \\
	p \left( y  | X_N \right) &= \text{Probability of } y  \text{ occurring given } X_N, \label{eq:condprob} \\
	  \qquad \qquad \; \; \beta_k &= k\text{th  coefficient}, k = 0,1,\ldots,N. \label{eq:betak} 
\end{align}
 In \eqref{eq:odds} and \eqref{eq:condprob}, the phrase ``given $X_{N}$''
 means \emph{given that a specific set of values of the variables} in $X_{N}$ occurred. The  logit \eqref{eq:gxlogit}  calculates the \textbf{log-odds}  of a  response   $ y $ occurring given that $X_N$ occurred. 
 
 I use the generic symbol $y$ in \eqref{eq:odds}--\eqref{eq:condprob} to match the usage in the  literature. Because $y$ has two possible values, I eliminate potential ambiguity in the meaning of $y$  by interpreting $y$ as representing a  \emph{specified response} or \emph{outcome}, which is either $y=1$ or $y=0$.  With this interpretation, the meaning of \eqref{eq:odds} is:   ``Odds of a specified  
 response (outcome) $y$ occurring given $X_N$,'' and likewise for \eqref{eq:condprob}.

The logit's coefficients $\beta_k$ are computed with a software package, of which  many are  available: My software package of choice is \textbf{R}. \cite{Rproject} Which  software package is used is immaterial. In all cases, calculating  the coefficients $ \beta_k$  requires specifying the values of the set of explanatory variables $X_N$ from measured  or  synthetic data or a combination thereof. 

The constant  $ \beta_0$ in \eqref{eq:gxlogit}  is commonly called  ``intercept''  in the literature because of its similarity to the term of the same name used in linear regression. The coefficient $ \beta_0$ is called ``intercept'' in \textbf{R} and in \cite{hosmer}.
%

\subsection{More About the Response and Explanatory Variables }      \label{subsec:thedata}
 
 In \eqref{eq:respvar} and \eqref{eq:expvar},  the response variable $y $ and the explanatory variable $ x_n$ are binary quantities, each toggling between two \textbf{states}  coded ``0'' and ``1'', respectively. It does not matter that  the  states of $y$ and $x_n$ have  identical values because $y$ and $x_n$ are \emph{categorical variables}, hence ``0'' and ``1'' are  \emph{labels}. The labels of response and explanatory variables are arbitrary and unrelated, hence their being identical is immaterial.  The software package \textbf{R} requires that categorical variables be explicitly identified as such.  For purposes of calculation,  the labels must be converted into numbers at some point.  Using $(0,1)$ to code the response and explanatory variables facilitates treating the labels as numbers as needed.\cite{hosmer} As explained in  \cite{hosmer},  certain formulas \emph{require} $(0,1)$ coding, else the formulas  have to be modified. All the formulas in this work require $(0,1)$ coding as specified in (\ref{eq:respvar}) and (\ref{eq:expvar}).
%

 \subsection{Defining the Events }      \label{subsec:events}
 
 \label{page:events}
 In subsequent sections I use a \emph{mathematical object} I call   \textbf{event} to obtain new  expressions for the odds ratio. Events and their properties are defined next. 
 
The  variables  $\{x_1,x_2,\ldots, x_N\}$ are said to be \textbf{realized} when all  $N$  variables   have specific  identities. For example,   a  realized set of three variables  might be $ \{x_1,x_2,x_3\}  \rightarrow   \{1,0,1\}$. Thus,  a realized set of  variables  consists of a sequence of zeros and ones.  comprise

 I define  \textbf{event}  as  \emph{a unique
  realization of the $N$ explanatory variables}.      For example,  there are four  events associated with the two-variable set $X_2=\{x_1,x_2\}$, namely,  $\big\{\{0,0\},\{0,1\},\{1,0\},\{1,1\}\big\}$. All events are realizations of the set $ X_{N}$  and vice versa. 
 
 The sequence of $N$ zeros and ones separated by a comma that makes up an event is equivalent to a \emph{positive $N$-digit binary integer}. For example, for the three-variable set $X_{3}$, the 3-digit binary integer $(010)_2$ corresponds to the event $\{0,1,0\}$. In all the events of the set $X_{N}$, the binary integers   correspond to the decimal integers $0,1,\ldots,2^N-1$.   \emph{Therefore, the number of distinct events is  equal to} $2^N$.
 
 The set of  $2^N$ events exhaustively comprises   all  possible  realizations of the set $X_N$.  \emph{Events are a property of the logit model and are completely defined by  the number of variables $N$}. Hence, \emph{events are independent of  data}.  Events  treat all the explanatory variables jointly, thereby \emph{contextualizing} the  variables.  
 
 %
 I use the following notation to denote  events.
\begin{align}
 	E_{\nu} &= \text{The } \nu\text{th event}, \label{eq:Enu} \\
	\nu &= \text{Event number}, \label{eq:nu} \\
	\nu &= 0,1,\ldots,2^N-1. \label{eq:nunumber} 
\end{align}

 A one-to-one correspondence exists between the  event number $\nu$ and  the event itself. Therefore, an event can be identified from  its number  and vice versa. 
As an example,  Table \ref{table:Ex1}  lists the $2^3 = 8$  events for $N=3$; specifying   $N=3$ sufficed to create the table.  The one-to-one correspondence between event numbers and events  is evident in the table.

\begin{table}[h!]
\centering
\begin{center}
\caption{Events  for the set of three  explanatory variables $X_3=\{x_1,x_2,x_3\}$.}\label{table:Ex1}
\begin{tabular}{c c c c }
\toprule
\ & \textbf{Decimal} &   &  \\
\textbf{Event} &\textbf{Event}   & \textbf{Binary} & \\
\textbf{Name}  &\textbf{Number} & \textbf{Event} & \textbf{Event}\\ 
 $\bm{(E_{\nu})}$  &$\bm{(\nu)}$& \textbf{Number}  &  $\bm{ \{x_1,x_2,x_3\}}$\\
\midrule
$E_0$ &0 & 000 & $\{0,0,0\}$ \\
$E_1$ & 1& 001 & $\{0,0,1\}$ \\
$E_2$ &2 & 010 & $\{0,1,0\}$ \\
$E_3$ &3& 011  & $\{0,1,1\}$\\
$E_4$ &4& 100  & $\{1,0,0\}$\\
 $E_5$ &5&101 & $\{1,0,1\}$ \\
$E_6$ & 6&110  & $\{1,1,0\}$\\
$E_7$ & 7&111  & $\{1,1,1\}$\\
 \bottomrule
\end{tabular}
\end{center}
\end{table}

Events similar to  $E_0=\{0,0,0\}$ and $E_7=\{1,1,1\}$ in Table \ref{table:Ex1}  have recurring roles in subsequent discussions and I have named them as follows:
\begin{align}
 	E_0 = \{0,0,\ldots,0\} & \rightarrow \text{\textbf{All-Zeros Event}},	\label{eq:allzeros} \\
	E_{2^N-1} =\{1,1,\ldots,1\} & \rightarrow \text{\textbf{All-Ones Event}}.	\label{eq:allones} 
\end{align}
%

%

 \section{The Basic Odds Ratio for Individual Variables}      \label{sec:oddsratio}

 \subsection{Defining the Odds Ratio}      \label{subsec:oddsratiodef}

The odds ratio statistic is well known, is widely used in many areas in addition to logistic regression, and needs little  introduction.\cite{oddsratio0,hosmer,szumilas}    Szumilas \cite{szumilas}  explains  and defines the odds ratio  particularly well:  

\label{page:ordef}
\begin{definition}[Odds Ratio \cite{szumilas}] 
``An odds ratio (OR) is a statistic that quantifies the strength of the association between two events,  \textsf{A} and  \textsf{B}. The odds ratio is defined as the ratio of the odds of  \textsf{A} in the \emph{presence} of \textsf{B} and the odds of \textsf{A} in the \emph{absence} of \textsf{B}.''  
\end{definition}.

The ``events''  \textsf{\textit{A}} and \textsf{\textit{B}} in this definition are unrelated to the events introduced in Section \ref{subsec:events} on~page~\pageref{page:events}. 

 \subsection{Interpreting the Odds Ratio}      \label{subsec:oddsratiointerpreted}
 
 Expressed schematically, the definition is  
\begin{equation}
	\text{Odds Ratio} = \frac{\text{Odds }(\textsf{\textit{A}} \text{ and } \textsf{\textit{B}} \text{ are present}) }{\text{Odds } (\textsf{\textit{A}} \text{ is present  and } \textsf{\textit{B}} \text{ is absent} )}. \label{eq:schematicOR} 
 \end{equation}

 It follows from \eqref{eq:schematicOR} that odds ratio dynamics  are  determined solely by \textsf{\textit{B}} because  \textsf{\textit{A}} does not change.
 
 Events \textsf{\textit{A}} and \textsf{\textit{B}} are either related or not. If not, then \textsf{\textit{B}} is immaterial and the ratio is \emph{equal to one}. If related, their connection  takes one of two  forms.   In one,  \textsf{\textit{B}}  makes the ratio  \emph{greater than one}. In the other, \textsf{\textit{B}}  makes the ratio  \emph{less than one}.  There are several ways for the odds ratio to be greater or less than one.
 
 The odds of outcomes associated with \textsf{\textit{A}}  are: (1) enhanced when the odds ratio is  greater than one, or (2) diminished when the odds ratio is less than one, or (3) unaffected when the odds ratio is equal to one.

 I explain the connection between the logit's variables $x_{n}$ and $\textsf{\textit{A}}$ and $\textsf{\textit{B}}$ on page \pageref{page:connection}.

 \subsection{Deriving the Basic Odds Ratio Formula}      \label{subsec:conventional}

In this section, I derive the known formula of the odds ratio of individual explanatory variables to establish a benchmark  for new results.   I shall call the benchmark odds ratio of individual variables  the  \textbf{basic odds ratio} and denote it   $\mathscr{B}(\cdot)$. 

To derive the basic odds ratio  formula, I  shall  follow the   algebraic approach  in the literature     (as in \cite{hosmer}, for example). The labels  0 and 1 will be treated as numbers.   I  shall derive the basic odds ratio for a  single-variable logit---the usual textbook choice---and for a two-variable logit.

To begin, consider    the following generic expression for the log-odds-ratio:
\begin{align}
 	\text{log-odds-ratio } &= \log(\text{numerator}/\text{denominator}),  \label{eq:genericOR1}  \\
	&= \log(\text{numerator}) - \log(\text{denominator}). \label{eq:genericOR2} 
\end{align}
  Thus, the  log-odds-ratio is  calculated for the variables   in the numerator \emph{relative} to the variables in the denominator.  Examples   in the  literature (\eg \cite{hosmer}) invariably use the All-Zeros Event \eqref{eq:allzeros}  in the denominator and I shall follow suit here. (But see the first remark in Section \ref{subsec:BORremarks}.) 

 The single-variable logit is  
\begin{align}
 	 \log \big( \mathscr{O}(X_1)  \big)   &= \beta_0 + \beta_1 x_1. \label{eq:logitx1}
\end{align}
Substituting $x_{1}=1$ and $x_{1} = 0$ in \eqref{eq:logitx1} yields 
\begin{align}
 \text{Numerator:} \quad	\log \big( \mathscr{O} (x_1=1) \big)  &= \beta_0 + \beta_1, \label{eq:numlogx1}\\
\text{Denominator:} \quad	\log \big( \mathscr{O} (x_1=0)  \big) &= \beta_0 \label{eq:denomlogx0}.
\end{align} 
Subtracting \eqref{eq:denomlogx0} from \eqref{eq:numlogx1} and substituting in \eqref{eq:genericOR1} yields 
\begin{align}
 	\log \bigg( \frac{\mathscr{O}(x_1=1)}{\mathscr{O}(x_1=0)} \bigg) = \beta_1. \label{eq:logORx1} 
\end{align}
Exponentiating \eqref{eq:logORx1} leads to the basic odds ratio of $x_{1}$, to wit,
\begin{align}
\mathscr{B}(x_1) &= \exp \Bigg(\log \bigg( \frac{\mathscr{O}(x_1=1)}{\mathscr{O}(x_1=0)} \bigg) \Bigg) = \exp(\beta_1), \label{eq:expbx1} 
\end{align}
 where ``$\exp(\cdot)$'' denotes the exponential function $\eu^{(\cdot)}$. The basic odds ratio  \eqref{eq:expbx1}  quantifies the odds ratio when  $x_1$  changes state  from 0 to 1. 
 
 Turning to coefficient $\beta_{0}$, \eqref{eq:denomlogx0}  shows that  $\beta_{0}$ is  the \emph{log-odds}---not the log-odds-\emph{ratio}---when all explanatory variables are equal to zero.   Hence,  $\exp(\beta_{0})$ pertains to \emph{odds} not to an \emph{odds ratio}; its treatment is explained in \cite{uclaidre2}. The numeric value of $\exp(\beta_0)$ is  a default output of \textbf{R} and possibly of other software packages used in logistic regression.

 Proceeding to the two-variable example,  its logit is
 \begin{align}
 	 \log \big( \mathscr{O}(X_2)  \big)   &= \beta_0 + \beta_1 x_1 + \beta_2 x_2. \label{eq:logitx1x2}
\end{align}

  I first obtain  the expression for  the basic odds ratio of  $x_1$, previously denoted $\mathscr{B}(x_1)$, then  derive $\mathscr{B}(x_2)$.   The evident pattern of these two  derivations  leads to the general benchmark basic odds ratio  formula, denoted $\mathscr{B}(x_n)$, applicable for $ n = 1,2,\ldots,N$.

 Moving on, observe that variables $x_1$ and $x_2$ combine in four ways, namely,   
\begin{align}
 x_1=0, & \quad x_2=0, 	\label{eq:00} 	\\
 x_1=0, & \quad x_2=1,	\label{eq:01} 	\\
 x_1=1, &\quad  x_2=0,	\label{eq:10} 	\\
 x_1=1, &\quad  x_2=1.	\label{eq:11} 
 \end{align}

 To define the numerator for $x_{1}$, notice that \eqref{eq:10}   is the only combination in which  variable $x_1$ has changed to state 1 from state 0 in \eqref{eq:00}   while  $x_2$ remains in state 0.     Substituting  \eqref{eq:10} and \eqref{eq:00} into \eqref{eq:logitx1x2} yields 
\begin{align}
 \text{Numerator:} \quad	\log \big( \mathscr{O} (x_1=1,x_2=0) \big)  &= \beta_0 + \beta_1, \label{eq:log10}\\
\text{Denominator:} \quad	\log \big( \mathscr{O} (x_1=0, x_2=0)  \big) &= \beta_0 \label{eq:log00}.
\end{align} 
Subtracting  \eqref{eq:log00}  from  \eqref{eq:log10} yields the log-odds-ratio 
\begin{align} 
\begin{split} \label{eq:logoddsbeta1}
 	 \log \big( \mathscr{O} (x_1=1, & \, x_2=0)  \big)  
	   -  \log \big( \mathscr{O} (x_1=0, x_2=0)  \big) \\
	 &  =  (\beta_0 + \beta_1) - \beta_0, \\
	 &  = \beta_1.
\end{split}
\end{align}
From \eqref{eq:logoddsbeta1},  the log-odds-ratio is thus equal to
\begin{align}
 	\log \bigg( \frac{\mathscr{O}(x_1=1, x_2=0)}{\mathscr{O}(x_1=0,x_2=0)} \bigg) &= \beta_1. \label{eq:logOR} 
\end{align}
Exponentiating \eqref{eq:logOR}   yields
\begin{align}
\exp \Bigg(\log \bigg( \frac{\mathscr{O}(x_1=1,x_2=0)}{\mathscr{O}(x_1=0,x_2=0)} \bigg) \Bigg)&= \exp(\beta_1), \label{eq:expb1} 
\end{align}
It follows from \eqref{eq:expb1} that the basic odds ratio of $x_1$ is 
\begin{align}
 	\mathscr{B}(x_1) =\frac{\mathscr{O}(x_1=1,x_2=0)}{\mathscr{O}(x_1=0,x_2=0)}  &= \exp( \beta_1), \label{eq:basicORx1} 
\end{align}
 which has the same \emph{form} as \eqref{eq:expbx1} as it should. In this case, however,  $\mathscr{B}(x_1)$ quantifies the consequence of  variable $x_1$ changing state from 0 to 1 while  $x_2$ remains unchanged in state 0.

The odds ratio $\mathscr{B}(x_2)$    is  derived in the same manner as $\mathscr{B}(x_1)$ but with \eqref{eq:01} in the numerator.     Substituting  \eqref{eq:01} and \eqref{eq:00} into \eqref{eq:logitx1x2} yields 
\begin{align}
 \text{Numerator:} \quad	\log \big( \mathscr{O} (x_1=0,x_2=1) \big)  &= \beta_0 + \beta_2, \label{eq:log01}\\
\text{Denominator:} \quad	\log \big( \mathscr{O} (x_1=0, x_2=0)  \big) &= \beta_0. \label{eq:log000} 
\end{align}
Subtracting \eqref{eq:log000} from \eqref{eq:log01} and simplifying yields
\begin{align}
 	\mathscr{B}(x_2) & =\frac{\mathscr{O}(x_1=0,x_2=1)}{\mathscr{O}(x_1=0,x_2=0)}  = \exp( \beta_2). \label{eq:basicORx2} 
\end{align}
The interpretation of  \eqref{eq:basicORx2} is similar to that of \eqref{eq:basicORx1} upon interchanging $x_{1}$ and $ x_{2}$.

Applying  the same   procedure to  a  logit with $N $  explanatory variables  yields the general expression for the basic odds ratio   of each individual variable $x_{n}$, namely,   
\begin{align}
 	\mathscr{B}(x_n) &= \frac{\mathscr{O}(x_1=0,x_2=0,\cdots, x_n=1, \cdots,x_N=0)}{\mathscr{O}(x_1=0,x_2=0,\cdots,x_n = 0, \cdots,x_N=0)} = \exp( \beta_n). \label{eq:BORfrac}    
\end{align}
Notice that only $x_{n}$ changes state; the other variables remain in state 0, thereby being effectively excluded from consideration, Written more compactly, the   \emph{benchmark} formula of the basic odds ratio is given by 
\begin{equation}
	\boxed{\mathscr{B}(x_n) = \exp( \beta_n), \quad n = 1,2,\ldots,N.} \label{eq:BOR}
 \end{equation} 

 The  basic-odds-ratio derivations show that all $\mathscr{B}(x_n)$  are independent: The derivation of each one  excludes all the others. 
  
 \subsection{Connection of Logit Variables with the Odds Ratio Definition}      \label{subsec:Connection}

\label{page:connection}  The  odds ratio was defined in Section \ref{subsec:oddsratiodef}  (page~\pageref{page:ordef}).  
Because \textsf{\textit{B}} is binary, one way to calculate the odds ratio is to use a single-variable logit such as  \eqref{eq:logitx1}.  The connection between  \textsf{\textit{B}} and  $x_{1}$ is straightforward. First, notice that states 0 and 1 can be interpreted as ``absent''  and   ``present,'' respectively. Thus,  (\ref{eq:expbx1}) shows that  $x_{1}$ is present in the numerator and absent in the denominator just as \textsf{\textit{B}}  is in \eqref{eq:schematicOR}.  Therefore, $x_{1}$ corresponds to  \textsf{\textit{B}}.

 \subsection{Remarks}      \label{subsec:BORremarks}

%
\begin{enumerate} \label{page:remarks} 
\item  To derive the general basic odds ratio $\mathscr{B}(x_n)$ (\eqref{eq:BORfrac}--\eqref{eq:BOR}), I  followed   common practice  in the  literature and used the All-Zeros Event \eqref{eq:allzeros} as the base.   Indeed, all the examples I found in the literature used the All-Zeros Event without explanation, which prompts the question: Is it necessary to use the All-Zeros Event exclusively? I show later that  it is not:  The same basic odds ratio can be obtained relative to other zero-one sequences. However, I will show that it  is often sufficient and  convenient  to use the All-Zeros Event.
\item In   \eqref{eq:BORfrac}, observe that only $x_n$ changes its state from 0 in the denominator to 1 in the numerator and  none of the other variables undergoes a change of state. It is the state-transition dynamic that matters. 
\item Notice that no odds ratio  was obtained  for  \eqref{eq:00},  which in this case is the All-Zeros Event for two variables. Whether an  odds ratio exists for an  All-Zeros Event will be addressed in Section \ref{subsec:inverseoddsratio}.
\end{enumerate}

 \section{Recasting  the Basic Odds Ratio}      \label{sec:reformulation}
 
 \subsection{The Basic Odds Ratio as a Function of Events}      \label{subsec:active}
 
 In this section, I  formulate the basic odds ratio  as a function of the events defined in Section \ref{subsec:events} (page \pageref{page:events}) and show that the new formulation yields an odds ratio formula that  matches the benchmark basic odds ratio   \eqref{eq:BOR}. The seeds of the events-based formulation are present in the derivation of \eqref{eq:BOR} but the  methodology used to obtain \eqref{eq:BOR}  is not conducive for those seeds to  germinate. The events-based formulation provides     insights about the odds ratio and leads to novel applications.     
 
 The following expression formulates the  basic odds ratio of  $x_n$  as a function of two events, a \textbf{reference event} denoted $E_r$, and a \textbf{target event} denoted $E_t$: 
 \begin{align}
 	B(x_n,E_r,E_t) & = \frac{ \mathscr{O}(E_t)  }{ \mathscr{O}(E_r) }, \textrm{where }  E_r \ne E_t, \text{  } n=1,2,\ldots,N,	\label{eq:ORrt}
%
\intertext{and}
%
E_r &= \textbf{\text{Reference Event}}, \label{eq:refevent} \\
E_t &= \textbf{\text{Target Event}}, \label{eq:tgtevent} \\
r,t &= \text{Event numbers}; r,t = 0,1,\ldots,2^N-1;  r\ne t, \label{eq:rt} \\ 
\mathscr{O}(E_r) &= \text{Odds of the reference event occurring}, \label{eq:oddsEt} \\
\mathscr{O}(E_t) &= \text{Odds of the target event occurring}. \label{eq:oddsEt}
\end{align}
I use the symbol  ``$B(\cdot)$'' in \eqref{eq:ORrt} to distinguish it from \eqref{eq:BOR} even though both yield the same result. 

Event $E_r$ is the base event in which  $x_n$ is in state 0, hence the name ``reference event.''  Event  $E_t$ is named  ``target event'' because it is the event in which $x_n$ has transitioned to state 1.  These  names  describe  the $0 \rightarrow 1$ state transition ($E_r \to E_t$ transition) of  $x_n$. Thus, the basic odds ratio $B(\cdot)$  is equal to the ratio of the odds of the target event relative to the odds of the reference event. As it should,    $ B(x_n,E_r,E_t)$ in \eqref{eq:ORrt}  has the same general form as \eqref{eq:BORfrac}. The conditions $E_r \ne E_t$ and $r \ne t$ are the same and  preclude $B(x_n,E_r,E_t) $ from being \emph{identically} equal to one. Nevertheless,  the odds ratio is equal to one when $E_r \ne E_t$ and there is no association between $E_r$  and $ E_t$.

 To use \eqref{eq:ORrt} to compute the basic odds ratio, I first rewrite  \eqref{eq:gxlogit}  as   
\begin{align}
 	 \log \big( \mathscr{O}(X_N) \big) & = \beta_0 +  \bm{\beta} \text{\textbullet} \bm{X}_N,	\label{eq:logit5}
\end{align}
where
\begin{align}
	\bm{\beta} &= [\beta_1,\beta_2,\ldots, \beta_N] = \text{Vector of } \beta \text{ coefficients}, \label{eq:betavector} \\
	 \bm{X}_N &= [x_1,x_2,\ldots,x_N],  \label{eq:xNvector}  \\
	 & = \text{Vector of  numerical explanatory variables}, \notag  \\
	 \bm{\beta} \text{\textbullet} \bm{X}_N &=  \text{Dot product of } \bm{\beta} \text{ and } \bm{X}_N. \label{eq:dotbetaX}  
\end{align}
The zeros and ones of the set $X_N$  are  treated as numbers in (\ref{eq:xNvector}).

Substituting  $E_r$ and $E_t$ for $\bm{X}_N$ in  \eqref{eq:logit5} yields
\begin{align}
 	 \log \big( \mathscr{O}(E_r) \big)  = \beta_0 +  \bm{\beta} \text{\textbullet} \bm{E}_r,	\label{eq:dotEr} \\
	 \log \big( \mathscr{O}(E_t) \big)  = \beta_0 +  \bm{\beta} \text{\textbullet} \bm{E}_t.	\label{eq:dotEt}
\end{align}
Exponentiating \eqref{eq:dotEr} and \eqref{eq:dotEt} leads to  
\begin{align}
 	\mathscr{O}(E_r)  &= \exp ( \beta_0 + \boldsymbol{\beta} \text{\textbullet} \bm{E_r}),	\label{eq:expEr} \\
	\mathscr{O}(E_t)  &= \exp ( \beta_0 + \boldsymbol{\beta} \text{\textbullet} \bm{E_t}).	\label{eq:expEt}
\end{align}
Then substituting  \eqref{eq:expEr} and \eqref{eq:expEt}  into \eqref{eq:ORrt} it follows that 
\begin{align}
 	B(x_n,E_r,E_t)  &= \frac{ \mathscr{O}(E_t)  }{ \mathscr{O}(E_r) } = \frac{ \exp \big(\beta_0 + \bm{\beta} \text{\textbullet} \bm{E}_t\big) }{\exp \big(\beta_0 + \bm{\beta} \text{\textbullet} \bm{E}_r\big)}, \label{eq:ORxnErEt1} \\
\intertext{and simplifying \eqref{eq:ORxnErEt1} yields}
	B(x_n,E_r,E_t) & = \frac{ \mathscr{O}(E_t)  }{ \mathscr{O}(E_r) }  = \exp \big( \beta_0 +  (\boldsymbol{\beta} \text{\textbullet} \bm{E}_t )  -\beta_0 -  (\boldsymbol{\beta} \text{\textbullet} \bm{E}_r ) \big),\label{eq:ORxnErEt2} 
\end{align}
whence it follows that  
\begin{align}
	\Aboxed{B(x_n,E_r,E_t) &= \exp \big(  \boldsymbol{\beta} \text{\textbullet} ( \bm{E}_t   -  \bm{E}_r ) \big).} \label{eq:GenODR}  
 \end{align} 
 When $E_{r}=E_{0}$,  \eqref{eq:GenODR} becomes simpler  as shown below.
 \begin{align}
	B(x_n,E_0,E_t) &= \exp \big(  \boldsymbol{\beta} \text{\textbullet} ( \bm{E}_t   -  \bm{E}_{0}) \big) =\exp \big(  \boldsymbol{\beta} \text{\textbullet} \bm{E}_t\big) \label{eq:GenODRE0}  
 \end{align}
 
 It remains to show that \eqref{eq:GenODR} reproduces the benchmark odds ratio formula \eqref{eq:BOR}. As explained next, reproducing the benchmark formula is achieved  by  selecting the appropriate events $E_r$ and $E_t$.
 
First, I show that \eqref{eq:GenODR} replicates \eqref{eq:basicORx1} and \eqref{eq:basicORx2} for  $N=2$. To do so, consider the  four events   defined in \eqref{eq:00}--\eqref{eq:11}, namely,   $E_0=\{0,0\}, E_1=\{0,1\},E_2=\{1,0\},E_3=\{1,1\}$. For variables $x_{1}$ and $x_{2}$, respectively substituting  $\bm{\beta}=[\beta_1,\beta_2]$ and the two event-pairs $(\bm{E}_r = \bm{E}_0 = [0,0],  \bm{E}_t = \bm{E}_2 = [1,0])$, and $(\bm{E}_r = \bm{E}_0 = [0,0], \bm{E}_t = \bm{E}_1=[0,1])$    in \eqref{eq:GenODR} yields  
\begin{align}
 	B(x_1,E_0,E_2) &= \exp \Big(  [\beta_1,\beta_2] \text{\textbullet} \big( [1,0]   -  [0,0]  \big) \Big)	= \exp(\beta_1), \label{eq:BE0E2} \\
	B(x_2,E_0,E_1) &= \exp \Big(  [\beta_1,\beta_2] \text{\textbullet} \big( [0,1]   -  [0,0] \big) \Big)	= \exp(\beta_2). \label{eq:BE0E1}
\end{align}
Thus \eqref{eq:BE0E2} and \eqref{eq:BE0E1} replicate \eqref{eq:basicORx1} and \eqref{eq:basicORx2}, respectively, which completes the  demonstration for $N=2$.

But \eqref{eq:GenODR} is more general than \eqref{eq:basicORx1} or \eqref{eq:basicORx2} and enables using other events to obtain the same result. For example,  substituting  $E_r=E_2=\{1,0\} \text{ and } E_t=E_3=\{1,1\}$ into \eqref{eq:GenODR} also yields $\exp(\beta_2)$ as shown below.
\begin{align}
 	 	B(x_2,E_2,E_3) &= \exp \Big(  [\beta_1,\beta_2] \text{\textbullet} \big( [1,1]   -  [1,0] ) \big) \Big)	= \exp (\beta_2). \label{eq:BE2E3}   
\end{align}
Thus,  \eqref{eq:BE2E3}  shows that it is not necessary to   use the All-Zeros Event  as reference to calculate the basic odds ratio.

Moreover, \eqref{eq:BE0E1} and \eqref{eq:BE2E3} show that the odds ratio $\exp(\beta_2)$ is the same regardless of the state of variable $x_1$, whose state is equal to 0 in \eqref{eq:BE0E1} and to 1 in \eqref{eq:BE2E3}. This demonstrates that the state of the ``other variable''---$x_1$ in this example---is \emph{immaterial provided it remains constant}. That the state of $x_1$---and in general the state of the invariant ``other variable(s)''---does not matter  demonstrates that the odds ratio of $x_2$ is \emph{context-free}. The \emph{context-free property of the basic odds ratio} is an important general property, and will be confirmed by other examples to be discussed.

Finally, I show that \eqref{eq:GenODR} reproduces the benchmark basic odds ratio $\mathscr{B}(x_n) $ for all $n$. To do this, set $E_r = E_0$ and $E_t = E_{2^{(N-n)}}$. The event $E_{2^{(N-n)}}$ singles out the individual variables $x_{n}$: Substituting  $n$ in the subscript $2^{(N-n)}$ identifies the event pertaining to  $x_n=1$. For example, for $N=2$, substituting  $n=1$ in  $2^{(2-n)}$  yields $E_{2^{(2-1)}}=E_2= \{1,0\}$, in which only $x_1 =1$, whereas substituting  $n=2$ yields    $E_{2^{(2-2)}}=E_1= \{0,1\}$, in which  only $x_{2} = 1$. In general,  
\begin{align}
 	B(x_n,E_0,E_{2^{(N-n)}}) & = \exp(\beta_{2^{(N-n)}}) = \mathscr{B}(x_n), \text{ where } n=1,2,\dots,N, \notag 
\end{align}
 which completes the demonstration.

 The following three examples illustrate the application of \eqref{eq:GenODR}. \\
 
 \textbf{Example 1: $\bm{N=1}$.}     

  This case has one variable, $x_1$, one coefficient, $\beta_1$, and  two events,  $E_0=\{0\}$ and $E_1=\{1\}$. Substituting $x_1, \beta_1$ and $E_r=E_0$ and $E_t=E_1$ in  \eqref{eq:GenODR} yields the expected odds ratio, namely, 
\begin{align}
 	B(x_1,E_0,E_1)    &= \exp( \beta_1) = \mathscr{B}(x_1). \notag 
\end{align}
The derivation of $\mathscr{B}(x_1)$ is thus readily obtained in terms of events $E_0$  and $E_1$. 

In the next two examples, the  events-based approach    yields new insights about the odds ratio of logits with more than one explanatory  variable. These  examples provide further confirmation  that: (1) although it is sufficient but not necessary to use  the All-Zeros Event  as reference, it is convenient and advantageous to use it, and (2) the basic odds ratio  is context-free.\\

 \textbf{Example 2: $\bm{N=2}$.}      
 
 Example 2  provides more evidence that there is more than one way to find the odds ratio of each variable and also of the  odds ratio's context-free property. Table \ref{table:Ex2}  lists the events for $N=2$ and Table \ref{table:Ex2Odr}  shows that the odds ratio of each variable can be computed in two ways.

\begin{table}[h!]
\caption{Example 2: Events  for the set of  two explanatory variables $X_2= \{x_1,x_2\}$.}\label{table:Ex2}
\begin{center}
\begin{tabular}{c c  }
\toprule
\textbf{Event}  & \\
\textbf{Name}  & \textbf{Event}\\ 
$\bm{(E_{\nu})}$  &  $\bm{ \{x_1,x_2\}}$\\
\midrule
$E_0$ & $\{0,0\}$ \\
$E_1$ & $\{0,1\}$ \\
$E_2$ & $\{1,0\}$ \\
$E_3$ & $\{1,1\}$\\
 \bottomrule
\end{tabular}
\end{center}
\end{table}
\begin{table}[h!]
\caption{Example 2: Odds Ratio for the set of two explanatory variables $X_2= \{x_1,x_2\}$.}\label{table:Ex2Odr}
\begin{center}
\begin{tabular}{c c c c c}
\toprule
  \textbf{Explanatory} & \textbf{Reference}  & \textbf{Target} & & \\
\textbf{Variable}  & \textbf{Event} &\textbf{Event} & & \textbf{Odds} \\ 
\small{ $\bm{(x_n)}$}  &  $\bm{ {(E_r) }}$ & $\bm{ (E_t) }$ &  $\bm{ {E_t -  E_r}}$ & \textbf{Ratio} \\
\toprule
$x_1$ & $E_0=\{0,0\}$   & $E_2=\{1,0\}$ & $[1,0]$ & $\exp ( \beta_1) $\\
$x_1$ & $E_1=\{0,1\}$ & $E_3=\{1,1\}$  & $[1,0]$& $\exp ( \beta_1) $\\
\midrule
$x_2$ & $E_0=\{0,0\}$  & $E_1=\{0,1\}$  & $[0,1]$ & $\exp ( \beta_2) $\\
$x_2$ & $E_2=\{1,0\}$  & $E_3=\{1,1\}$  & $[0,1]$ &$\exp ( \beta_2) $\\
 \bottomrule
\end{tabular}
\end{center}
\end{table}

  Table \ref{table:Ex2Odr} shows that there are two  $(E_r, E_t)$ pairs for each variable. For the variable $x_1$, the event-pairs are $(E_r,E_t) \rightarrow (E_0,E_2)$ and $(E_r,E_t) \rightarrow (E_1,E_3)$. In both,  $x_1$ changes its state from 0 in $E_r$  to 1 in $E_t$  while the state of  $x_2$ remains constant at either 0 or 1. Thus the odds ratio of $x_1$  is the same regardless of the state of $x_2$. A similar phenomenon is evident for $x_2$. This demonstrates the context-free property of  the odds ratio  of a single variable:   the state of the  other variable does not matter provided it does not change.
  
   Table \ref{table:Ex2Odr} shows that for each variable,  it suffices to  use only one of the two $(E_r,E_t)$ pairs to compute the odds ratio and that event $E_0$ is always one of the two reference events.  Therefore, although it is not necessary to  use  the All-Zeros Event as reference, its  ubiquity  makes its use convenient for all odds ratio calculations. \emph{Thus, the All-Zeros Event can be considered a universal reference for all individual variables}. \\

 \textbf{Example 3: $\bm{N=3}$.}     
 This example offers further confirmation that  the basic odds ratio of each variable can be obtained in more than one way and also of the basic odds ratio's general context-free property.   
\begin{table}[h!]
\caption{Example 3: Events  for the set of three  explanatory variables $X_3=\{x_1,x_2,x_3\}$.}
\begin{center}
\begin{tabular}{c c  } 
\toprule
\textbf{\small{Event}} & \\
\textbf{\small{Name}}   & \textbf{\small{Event}}\\ 
\small{ $\bm{(E_{\nu})}$}   & \small{ $\bm{ \{x_1,x_2,x_3\}}$}\\
\midrule
$E_0$ & $\{0,0,0\}$ \\
$E_1$ &  $\{0,0,1\}$ \\
$E_2$ & $\{0,1,0\}$ \\
$E_3$ & $\{0,1,1\}$\\
$E_4$ & $\{1,0,0\}$\\
 $E_5$ & $\{1,0,1\}$ \\
$E_6$ &  $\{1,1,0\}$\\
$E_7$ &  $\{1,1,1\}$\\
 \bottomrule
\end{tabular}
\end{center}
\label{table:Ex3}
\end{table}
\begin{table}[h!]
\caption{Example 3: Odds Ratio for the three individual explanatory variables $x_1, x_2, x_3$.}
\begin{center}
\begin{tabular}{c c c c c }
\toprule
  \small{\textbf{Explanatory}} & \textbf{\small{Reference}}  & \textbf{\small{Target}} & & \\
\textbf{\small{Variable}}  & \textbf{\small{Event}} &\textbf{\small{Event}} & & \small{\textbf{Odds}} \\ 
\small{ $\bm{(x_n)}$}  & \small{$\bm{ {(E_r)}}$} & \small{$\bm{(E_t)}$} & \small{$\bm{E_t -  E_r}$} & \small{\textbf{Ratio}} \\
\toprule
\small{$x_1$} & \small{$E_0=\{0,0,0\}$}   &\small{$E_4=\{1,0,0\}$} & \small{$[1,0,0]$} & \small{$\exp (\beta_1) $}\\
\small{$x_1$} & \small{$E_1=\{0,0,1\}$}  &\small{$E_5=\{1,0,1\}$} & \small{$[1,0,0]$} & \small{$\exp (\beta_1) $}\\
\small{$x_1$} & \small{$E_2=\{0,1,0\}$}  & \small{$E_6=\{1,1,0\}$} & \small{$[1,0,0]$} & \small{$\exp (\beta_1) $}\\
\small{$x_1$} & \small{$E_3=\{0,1,1\}$} & \small{$E_7=\{1,1,1\}$} & \small{$[1,0,0]$}& \small{$\exp (\beta_1) $}\\
\midrule
    \small{$x_2$} &\small{$E_0=\{0,0,0\}$}  & \small{$E_2=\{0,1,0\}$} & \small{$[0,1,0]$} & \small{$\exp (\beta_2) $}\\
\small{$x_2$} & \small{$E_1=\{0,0,1\}$} & \small{$E_3=\{0,1,1\}$} & \small{$[0,1,0]$} &  \small{$\exp (\beta_2) $} \\
    \small{$x_2$} & \small{ $E_4=\{1,0,0\}$ } & \small{$E_6=\{1,1,0\}$} & \small{$[0,1,0]$} & \small{$\exp (\beta_2) $ }\\
    \small{$x_2$} & \small{$E_5=\{1,0,1\}$}  & \small{$E_7=\{1,1,1\}$} & \small{$[0,1,0]$} & \small{$\exp (\beta_2)$} \\
\midrule
\small{$x_3$} & \small{$E_0=\{0,0,0\}$}  & \small{$E_1=\{0,0,1\}$}  & \small{$[0,0,1]$} & \small{$\exp (\beta_3)$}\\
\small{$x_3$} & \small{$E_2=\{0,1,0\}$}  & \small{$E_3=\{0,1,1\}$} & \small{$[0,0,1]$} & \small{$\exp (\beta_3)$}\\
\small{$x_3$} & \small{$E_4=\{1,0,0\}$}  & \small{$E_5=\{1,0,1\}$} & \small{$[0,0,1]$} &$\exp (\beta_3)$\\
\small{$x_3$} & \small{$E_6=\{1,1,0\}$}  & \small{$E_7=\{1,1,1\}$} & \small{$[0,0,1]$} & \small{$\exp (\beta_3)$}\\
 \bottomrule
\end{tabular}
\end{center}
\label{table:Ex3Odr}
\end{table}

  Table \ref{table:Ex3} lists the events for the set $X_{3}$ and Table \ref{table:Ex3Odr}   shows that there are four $(E_r, E_t)$ pairs  for each individual variable and all  produce the same odds ratio.  Table \ref{table:Ex3Odr}  provides further evidence of the context-free property of  the odds ratio of individual variables. Observe that $E_r$ and $E_t$ are identical except for the variable whose  state changes from 0 in the reference event to 1 in the target event. 
  
  As in Table \ref{table:Ex2Odr},  Table \ref{table:Ex3Odr} shows that event $E_0$ is ubiquitous  as a reference for all basic odds ratios, thereby affirming the convenience and universality of  the All-Zeros Event  as  the reference event for all basic odds ratio calculations.

 \subsection{Attributes of the Basic Odds Ratio }      \label{subsec:attributes}
The basic odds ratio and its event-based formula \eqref{eq:GenODR}  have the following attributes:
\begin{enumerate}
\item The basic odds ratio quantifies the effect on the odds induced when the state of one individual variable transitions from 0 to 1 while the state of the other variables remains constant at either 0 or 1.
\item  The basic odds ratio is context-free: The state of the other variables does not matter---their respective  states can be either 0 or 1---as long as the states remain constant in the transition $E_{r} \rightarrow E_{t}$. Thus, \emph{the basic odds ratio has the property that it separates one individual variable from the influence of the other variables.}
\item The reference and target events $E_r$ and $E_t$ differ only in that the state of the individual variable of interest is equal to 0 in $E_r$ and to 1 in $E_t$.
\item Several variable combinations  can be used to calculate the basic odds ratio. The number of  combinations increases with $N$.
\item It is both sufficient and convenient to  always use the All-Zeros Event $E_0$  as the reference event. 

\end{enumerate}

 \section{The Group Odds Ratio: A Generalized \\ Multivariable Odds Ratio}      \label{sec:groupoddsratio}
 
\subsection{Deriving the Group Odds Ratio Formula}      \label{subsec:groupoddsratio}
 Thus far, I have dealt with the odds ratio of individual explanatory variables---standard fare in logistic regression. Such  variables change state  one variable at a time: the basic odds ratio  \eqref{eq:GenODR}  quantifies the effect of that single-variable state change. But what is the effect when a \emph{subset of several variables} changes state from 0 to 1 simultaneously? To answer this question requires computing a different odds ratio, a \emph{generalized multivariable odds ratio}, which I have named \textbf{Group Odds Ratio}. The Group Odds Ratio  is calculated using a generalized version of \eqref{eq:GenODR}, namely,    
\begin{align}
 	\Aboxed{G (S_k,E_r,E_t) &= \frac{ \mathscr{O}(E_t)}{ \mathscr{O}(E_r)} = \exp  \big(  \boldsymbol{\beta} \text{\textbullet} (  \bm{E}_t - \bm{E}_r) \big).}	\label{eq:Gxrt}  
\end{align}
In  \eqref{eq:Gxrt}, the terms $E_r,E_t, \boldsymbol{\beta}, \bm{E}_t, \bm{E}_r $ are the same as in \eqref{eq:GenODR}; the term $S_k$ will be defined  shortly.

As shown in \eqref{eq:Gxrt}, the Group Odds Ratio consists of the exponential function whose exponent is a sum of beta coefficients. This is superficially the same as in  \eqref{eq:GenODR}, but in \eqref{eq:GenODR}  the sum always lands on a single variable: All but one of the summands  cancel out.  In contrast, in \eqref{eq:Gxrt} the sum of coefficients may have more than one component. For example, suppose that $N=3$ and $\bm{\beta} = [ \beta_1,\beta_2,\beta_3]$. Ignoring $S_k$ for the moment, suppose also  that $\bm{E}_t - \bm{E}_r=[1,0,1]$. Then the Group Odds Ratio is equal to $\exp \big([ \beta_1,\beta_2,\beta_3] \text{\textbullet} [1,0,1] \big)=\exp\big(\beta_1+\beta_3\big)$. But $\exp\big(\beta_1+\beta_3\big) = \exp(\beta_1) \times \exp(\beta_3) $, which  is the product of the basic odds ratios $\exp(\beta_1)$ and  $\exp(\beta_3)$. Thus, the Group Odds Ratio can be expressed in two equivalent ways: (1) as the exponential function raised to a power that is equal to the sum of a subset of beta coefficients, or (2) as the product of the same subset of basic odds ratios. Either as a sum or as a product, and in keeping with their foundational property, \emph{the basic odds ratios are the building blocks of the Group Odds Ratios}.

The generalization of \eqref{eq:GenODR}  consists of replacing the  single  variable $x_n$  in \eqref{eq:GenODR}   with $S_k$  in \eqref{eq:Gxrt}, where $S_k$ is a set defined as 
\begin{align}
 	S_k &=  k\text{th subset  of the \textbf{power set} of  the set } X_N.	\label{eq:Sk} 
\end{align}
 The \emph{power set} is the \emph{set of all subsets} of  the set $X_N$ and is denoted $\mathcal{P}(X_N)$. \cite{MacLane, powerset} The power set  $\mathcal{P}(X_N)$  contains $2^N$ subsets, including  the \emph{empty set}, denoted $\emptyset$, and  the set $X_N$ itself. I numbered the subsets  $S_k$ the same way as the events, namely, $k=0,1,\ldots,2^N-1$, where $k=0$ corresponds to the empty set. For example, the power set of the set  $X_2=\{x_1,x_2\}$ has $2^2=4$ subsets numbered $k = 0,1,2,3$. The power set itself is $\mathcal{P}(X_2)=\big\{S_0=\emptyset,S_1=\{x_1\},S_2=\{x_2\},S_3=\{x_1,x_2\}\big\}$. I will show later that for  $k = 1,2,\ldots,2^{N}-1$ there is a one-to-one correspondence between the subsets  $S_k$  and the target events $E_t$.

 Changing $x_n$ in \eqref{eq:GenODR} to $S_k$ in \eqref{eq:Gxrt}---a seemingly minor modification---makes all the difference because in \eqref{eq:Gxrt}  the subsets $S_k$  encompass all $2^{N}$  combinations of the variables $x_n$, thereby enabling calculating the  odds ratio when  several variables  change state from 0 to 1 simultaneously. Therefore, \eqref{eq:Gxrt} \emph{contextualizes} the odd ratios of the multivariate subsets. Because the power set $\mathcal{P}(X_N)$ includes a subset for each of the $N$ individual variables, it follows that    \eqref{eq:Gxrt} 
 \emph{subsumes} \eqref{eq:GenODR}.  Thus, \eqref{eq:Gxrt}  suffices for computing all the odds ratios, be they  basic odds ratios or Group Odds Ratios.
 For example, Table \ref{table:subsetsNeq3} lists all  $2^{3}=8$ subsets of the power set $\mathcal{P}(X_3)$ and shows that the single-variable subsets $S_1=\{x_{3}\},S_2=\{x_{2}\},S_4=\{x_{1}\}$ are included. The table shows that there are four multivariable subsets.
\begin{table}[h!]
\caption{The eight subsets of   the power set $\mathcal{P}(X_3)$ of the set $X_3=\{x_1,x_2,x_3\}$.}
\begin{center} 
\begin{tabular}{c c  } 
\toprule
  \textbf{Subset Name} & \\
  $\bm{(S_k)}$ &  \textbf{Subset}\\
\midrule
$S_0$ & $\emptyset$  \\
$S_1$ & $\{x_3\}$  \\
$S_2 $& $\{x_2\}$ \\
$S_3$ &  $\{x_2,x_3\}$ \\
$S_4$ & $\{x_1\}$  \\
$S_5$ & $\{x_1,x_3\}$ \\
$S_6$ & $\{x_1,x_2\}$ \\
$S_7$ & $\{x_1,x_2,x_3\}$ \\
 \bottomrule
\end{tabular}
\end{center}
\label{table:subsetsNeq3}
\end{table}

Having defined $S_k$, $E_r$ and $E_t$, I show next  how they are actualized, starting with the events $E_r$ and $E_t$.  
  Events $E_r$ and $E_t$ are identical except for the subset of variables that change from state 0  in $E_r$ to state 1 in $E_t$.  
  
 As an example, Table \ref{table:GroupNeq3} lists the reference and target events of the four  multivariable subsets in Table \ref{table:subsetsNeq3}. The first row of Table \ref{table:GroupNeq3} shows that for  subset $S_3=\{x_{2},x_{3}\}$, $x_{2}$ and $x_{3}$ change state from 0 in event $E_0$ to 1 in event $ E_3$ while $x_1$ remains in state 0. Therefore, $E_{3} - E_{0}=E_{3}$.
 Moreover, Table \ref{table:GroupNeq3} shows that for the two-variable subsets $S_3,S_5,S_6$, the same Group Odds Ratio  can  be computed for the same  subset using two  $(E_r, E_t)$ pairs.  This is akin to the situation  for individual  variables, for which Table \ref{table:Ex3Odr} showed for each variable that four $(E_r, E_t)$ pairs can be used to compute the same basic odds ratio.  
 Table \ref{table:GroupNeq3} also shows that for subset $S_7$---the All-Ones Event---only the  $(E_7, E_0)$ pair can be used to compute the Group Odds Ratio.  
 
 The results just described are general:  Multiple $(E_r, E_t)$ pairs can be used to compute the Group Odds Ratio  for all subsets containing at least one but fewer than $N$ variables (subsets $S_k$, where $k=1,2,\ldots,2^N-2$) and only the pair $(E_0,E_{2^N-1})$, can be used to compute the Group Odds Ratio for all $N$ variables (subset $S_{2^N-1}$).
\begin{table}[h!]
\caption{Reference and target events  for calculating the Group Odds Ratio  of the multivariate subsets of set $X_3$.}
\begin{center}
\begin{tabular}{c c c c  }
\toprule
  \textbf{Multivariate} & \textbf{Target} & \textbf{Reference} &  \\
  \textbf{Subsets} & \textbf{Event} &\textbf{Event} &  \\ 
  $ \bm{S_{3},S_{5},S_{6},S_{7}}$  & $\bm{ {(E_t) }}$ & $\bm{ (E_r) }$ &  $\bm{ {E_t -  E_r}}$  \\
\midrule
$S_3=\{x_2,x_3\}$ & $E_3=\{0,1,1\}$  & $E_0=\{0,0,0\}$ & $[0,1,1]$ \\
$S_3=\{x_2,x_3\}$ & $E_7=\{1,1,1\}$ & $E_4=\{1,0,0\}$ & $[0,1,1]$ \\
\midrule 
$S_5=\{x_1,x3\}$ & $E_5=\{1,0,1\}$ & $E_0=\{0,0,0\}$ & $[1,0,1]$\\
$S_5=\{x_1,x3\}$ & $E_7=\{1,1,1\}$ & $E_2=\{0,1,0\}$ & $[1,0,1]$\\
\midrule
$S_6=\{x_1,x_2\}$ & $E_6=\{1,1,0\}$ & $E_0=\{0,0,0\}$ & $[1,1,0]$ \\
$S_6=\{x_1,x_2\}$ & $E_7=\{1,1,1\}$ & $E_1=\{0,0,1\}$ & $[1,1,0]$ \\
\midrule
$S_7=\{x_1,x_2,x_3\}$ & $E_7=\{1,1,1\}$ & $E_0=\{0,0,0\}$ & $[1,1,1]$ \\
 \bottomrule
\end{tabular}
\end{center}
\label{table:GroupNeq3}
\end{table}

Tables \ref{table:Ex3Odr} and \ref{table:GroupNeq3} show that event $E_0$ appears as a reference event for all the subsets for both single variables (Table \ref{table:Ex3Odr}) and  multiple variables (Table \ref{table:GroupNeq3}).  Table \ref{table:GroupNeq3}  also shows that the All-Zeros Event is the \emph{only}  event that can be used as the reference   when the subset corresponds to the All-Ones Event, which is event $E_{7}$ in the example and event $E_{2^{N}-1}$ in general.  Thus, to include subset $S_{2^N-1}$ together with all the other subsets,  it is \emph{necessary} to use $E_0$  as the reference event for \emph{all} the subsets, which  settles the universal-reference status of the All-Zeros Event. (To no one's surprise, the literature got it right; it just does  not explain the reason.) Table \ref{table:GroupNeq3} also shows that with $E_0$ as reference, the subscripts of the subsets match the subscripts of the target events. Hence, for the target events in \eqref{eq:GX0t}, the event numbers are $t=1,2,\dots,2^N-1$. Substituting $E_0$ for $E_r$ and changing the subset's subscript  from $k$ to $t$    simplifies \eqref{eq:Gxrt} as follows:
\begin{align}
 	G ( S_t,E_0, E_t) &= \frac{ \mathscr{O}(E_t)}{ \mathscr{O}(E_0)} = \exp \big( \boldsymbol{\beta} \text{\textbullet} (\bm{E}_t-\bm{E}_0)\big), \label{eq:GXSt} 
\end{align}
\begin{align}
	\Aboxed{G ( S_t,E_0, E_t) &= \frac{ \mathscr{O}(E_t)}{ \mathscr{O}(E_0)}  =\exp( \boldsymbol{\beta} \text{\textbullet} \bm{E}_t).} \label{eq:GX0t}
\end{align}

The  composition of the subsets is thus determined by the target events.  Specifically,  the subsets are specified by
\begin{align}
 	S_t &= \big\{ \bm{E}_t \odot \bm{X}_N \big\}  \backslash \{0\},	\label{eq:subsetHadamard} 
\end{align}
where $t=1,\ldots,2^N-1$, the symbol ``$ \odot $'' denotes the \textbf{Hadamard Product} \cite{hadamard}, and ``$\backslash \{0\}$'' means that the zeros have been removed from the product. The subset $S_t$ thus consists of the zero-free Hadamard  product  of the vectors $\bm{E_{t}}$ and $\bm{X_{N}}$. Therefore, \eqref{eq:subsetHadamard} establishes the one-to-one correspondence between the subsets and the target events. In  \eqref{eq:subsetHadamard}, the terms $\bm{E}_t$ and $\bm{X}_N$ mean that the sets $E_t$ and $X_N$ are treated as numerical vectors. The curly braces that enclose the Hadamard product mean that the resulting product reverts to a set. The symbol ``$\backslash \{0\}$'' means that the  set has no zeros.
 The close connection between events, subsets, and odds ratios  shows that the \emph{events underlie  everything}.  
\begin{table}[h!]
\caption{Events, subsets, and Group Odds Ratios for the set $X_3=\{x_1,x_2,x_3\}$.}
\begin{center}
\begin{tabular}{c c c   }
\toprule
 \textbf{Target}  & \textbf{Variable}   & \textbf{Group}   \\
\textbf{Event}  & \textbf{Subset} &\textbf{Odds Ratio}  \\ 
 $\bm{ {(E_t) }$} & $\bm{(S_t)}$ &  $\bm{\big( G( S_t,E_0,E_t)\big) }  $    \\
\midrule
$E_1=\{0,0,1\}$ & $S_1=\{x_3\}$   & $\exp(\beta_3)$  \\
$E_2=\{0,1,0\}$ & $S_2=\{x_2\}$   & $\exp(\beta_2)$  \\
 $E_3=\{0,1,1\}$ & $S_3=\{x_2,x_3\}$   & $\exp(\beta_2 + \beta_3)$   \\
 $E_4=\{1,0,0\}$ & $S_4=\{x_1\}$   & $\exp(\beta_1)$  \\
$E_5=\{1,0,1\}$ &$S_5=\{x_1,x_3\}$   & $\exp(\beta_1 + \beta_3)$ \\
$E_6=\{1,1,0\}$ & $S_6=\{x_1,x_2\}$  & $\exp(\beta_1 + \beta_2)$ \\
$E_7=\{1,1,1\}$ & $S_7=\{x_1,x_2,x_3\}$  & $\exp(\beta_1+\beta_2 + \beta_3)$  \\
 \bottomrule
\end{tabular}
\end{center}
\label{table:GroupODRNeq3}
\end{table}

Table \ref{table:GroupODRNeq3}  lists all the events, subsets, and group odds ratios  for $N=3$, with  $E_0$ as  reference event.    
 The table  exhibits the one-to-one correspondence between  $S_t$ and $E_t$.
For example, for $E_3=\{0,1,1\}$ and $X_3=\{x_1,x_2,x_3\}$, the Hadamard product is $ \bm{E}_3 \odot \bm{X}_3 = [0,1,1] \odot [x_1,x_2,x_3] =[0,x_2,x_3] \rightarrow \{0,x_2,x_3\}$, which becomes   $S_3 = \{x_2,x_3\}$  upon deleting 0.

 Table \ref{table:GroupODRNeq3} confirms that the odds ratio of individual variables is subsumed by  \eqref{eq:GX0t} and that the exponents of the odds ratios of multivariable subsets consist of a sum of $\beta$ coefficients. As noted previously, the multivariable odds ratios  are products of the basic odds ratios. For example, using \eqref{eq:GX0t} with   $t=5$ 
\begin{align*}
 	G(S_5,E_0,E_5) &=G\big(\{x_1,x_3\},\{0,0,0\},\{1,0,1\}\big) \\
	&= \exp\big([\beta_1,\beta_2,\beta_3]  \text{\textbullet} [1,0,1]\big) \\
	&= \exp(\beta_1+ \beta_3 ) = \exp(\beta_1) \times \exp( \beta_3 ).
\end{align*}
%
%

\subsection{Inverse Odds Ratio}      \label{subsec:inverseoddsratio}

In the discussion of the two-variable example  (page \pageref{page:remarks}), I noted that an odds ratio was not calculated for  $\{x_1=0,  x_2=0\}$ (see \eqref{eq:00}), by now recognized as a two-variable All-Zeros Event.  I noted further that an odds ratio exists for the All-Zeros Event.   In fact, such an inverse odds ratio is implicit in the definition (page~\pageref{page:ordef}) by interchanging ``present'' and ``absent.'' The interchange reverses the odds ratio's inequality while retaining the equality (if any) and maintaining the correct interpretation of their respective meanings.

 Computationally,  the All-Zeros Event  has an inverse obtained by interchanging the target and reference events and using the All-Ones-Event as reference. Specifically, the inverse is 
%
%
\begin{align}
\begin{split} \label{eq:GORInverse}
 	G ( S_{2^N-1} ,E_{2^N-1}, E_0) &= \frac{ \mathscr{O}(E_0)}{ \mathscr{O}(E_{2^N-1})} =  \exp( \boldsymbol{- \beta} \text{\textbullet} \bm{E}_{2^N-1}), \\
	&=  \exp\big(-(\beta_1+\beta_2+\cdots+\beta_{2^N-1})\big). 
\end{split}
\end{align}

 In \eqref{eq:GORInverse}, $E_{0}$, the All-Zeros Event, is now the \emph{endpoint} for the \emph{simultaneous} transition of all the variables  from state 1 to state 0. Therefore, $E_{2^N-1}$, the All-Ones-Event,  must be   the reference for such a state transition.

------------------------

\section{Completing the Story}      \label{sec:story}

\subsection{Prologue}      \label{subsec:prelims}

 Every data set contains  a  story waiting to be told. The story is often elicited from the data via statistical  analysis. For the binary type of data considered in this note, the analysis may use  \emph{logistic regression}, which endeavors to model the data and whose output is an \emph{odds ratio} for each explanatory variable: I call such odds ratios \textit{basic odds ratios}.  This section shows how  basic odds ratios come together to tell the story.
\subsection{Why the Group Odds Ratio?}      \label{subsec:whygroup}

Although the basic odds ratio is the fundamental statistic of logistic regression, in a multivariable setting  basic odds ratios provide only partial information because they lack context. The basic odds ratio  is exactly what it is designed for, namely,  a context-free statistic for a single variable. The ability to isolate the effect of one variable from the combined effects of all other variables is powerful but presages a weakness: The basic odds ratio acts as if only one variable  exists. But the other variables do exist. In fact, \emph{all} the variables are always \emph{realized jointly},  as  described by the events (for example, as illustrated in Table \ref{table:Ex3}). The  basic odds ratio's contextual shortcomings become glaring when the logit model has three or more variables. 

In contrast to the basic odds ratio, the Group Odds Ratio provides contextual   information for subsets consisting of all the combinations of two or more variables, ending with the All-Ones-Event (see, for example, subsets $S_3,S_5,S_6,S_7$  in Table \ref{table:GroupODRNeq3}).  Only the All-Ones-Event   provides full  contextual information. Subsets containing at least two  but fewer than $N$ variables provide partially-contextual information; the odds ratio for those subsets is context-free with respect to the variables not in the subset. Contextual information increases as the subsets add variables. 

Finally, the ensemble of Group Odds Ratios provides a framework for analyzing  odds-ratio dynamics as the odds ratios vary across events.  The ensemble of Group Odds Ratios (such as  displayed in Table \ref{table:GroupODRNeq3})   fully describes those dynamics. A plot of Group Odds Ratios would show the dynamics at a glance and  enable visual identification of maxima and minima. Moreover, statistical analysis of the ensemble can yield  insights about the  odds ratios 

\section{Conclusions}      \label{sec:conclusions}

The  odds ratio of    \emph{individual} explanatory variables---named ``basic odds ratio''  in this work---is the principal  statistic of logistic regression. The basic odds ratio is foundational  and purposefully devoid of context. It is a powerful statistic that separates   the effect of one explanatory variable from the influence of  all other variables. 
 
 The basic odds ratio    answers the   question:
 What is the effect  on the odds of a desired response occurring    when the state of  \emph{exactly one}  explanatory variable  changes from state 0  to state 1   while the respective states of the other     variables  remain unchanged? 
 
The purposeful lack of context limits the scope of each basic odds ratio  to  one  explanatory variable. But in multivariate logistic regression all the  variables always appear jointly in various combinations about which basic odds ratios provide no information. For this reason, basic odds ratios do not tell the full story of the odds ratios. 

The complete  odds-ratio story answers a more general question that goes beyond the basic odds ratio while encompassing  it, namely: What is the effect  on the odds of a desired response  occurring    when the state of a \emph{subset} consisting of more than one  explanatory variable changes \emph{simultaneously} from state 0  to state 1   while the respective states of the variables not in the subset remain unchanged? 

To answer the preceding question and complete the odds-ratio story requires contextual knowledge obtained by  treating all the   variables simultaneously.  The generalized multivariable odds-ratio  statistic named ``Group Odds Ratio''  derived in this work     provides both the requisite contextual information and the computational methodology to calculate it.  The complete odds-ratios story is told by the ensemble of Group Odds Ratios, which contains all the odds ratios including the basic odds ratios.


%

\end{document}